\documentclass[floats, aps, showpacs, twocolumn, superscriptaddress, prl,
    reprint, floatfix]{revtex4-1}

\usepackage{graphicx}
\usepackage{amsmath, amssymb, amstext, amsthm, amsfonts}

\usepackage{times}

\begin{document}

\title{Hierarchical Fractal Weyl Laws for Chaotic Resonance States in Open
Mixed Systems}

\author{M.\,J.\ K\"orber}
\affiliation{Technische Universit\"at Dresden, 
             Institut f\"ur Theoretische Physik and Center for Dynamics,
             01062 Dresden, Germany}
\affiliation{Max-Planck-Institut f\"ur Physik komplexer Systeme, N\"othnitzer
Stra\ss{}e 38, 01187 Dresden, Germany}

\author{M.\ Michler}
\affiliation{Technische Universit\"at Dresden, 
             Institut f\"ur Theoretische Physik and Center for Dynamics,
             01062 Dresden, Germany}

\author{A.\ B\"acker}
\affiliation{Technische Universit\"at Dresden, 
             Institut f\"ur Theoretische Physik and Center for Dynamics,
             01062 Dresden, Germany}
\affiliation{Max-Planck-Institut f\"ur Physik komplexer Systeme, N\"othnitzer
Stra\ss{}e 38, 01187 Dresden, Germany}

\author{R.\ Ketzmerick}
\affiliation{Technische Universit\"at Dresden, 
             Institut f\"ur Theoretische Physik and Center for Dynamics,
             01062 Dresden, Germany}
\affiliation{Max-Planck-Institut f\"ur Physik komplexer Systeme, N\"othnitzer
Stra\ss{}e 38, 01187 Dresden, Germany}

\date{September 16, 2013}

\begin{abstract}
In open chaotic systems the number of long-lived resonance states obeys a
fractal Weyl law, which depends on the fractal dimension of the chaotic
saddle. We study the generic case of a mixed phase space with regular and
chaotic dynamics. We find a hierarchy of fractal Weyl laws, one for each
region of the hierarchical decomposition of the chaotic phase-space component.
This is based on our observation of hierarchical resonance states localizing on
these regions. Numerically this is verified for the standard map and a
hierarchical
model system.
\end{abstract}

\maketitle

\noindent


It is just a century ago that Hermann Weyl published his celebrated theorem on 
the asymptotic distribution of eigenmodes of the Helmholtz equation in a
bounded domain \cite{Wey12} which has found fundamental applications in the
context of acoustics, optical cavities and quantum billiards \cite{Kac66,
Sto99, AreNitPetSte2009}. 
For a quantum billiard with a $d$ dimensional phase space the number
$\mathcal{N}(k)$ of eigenmodes with a wave number below $k$ is on average and
in the limit of large $k$ given by $\mathcal{N} (k) \sim k^{d/2}$ up to
corrections of higher order \cite{BalBlo1970, BalHil1976, BerHow1994,
Ivr2010, BaeKetLoeSch2011}. Only recently, this fundamental question has been
addressed for open scattering systems, where for the case of fully chaotic
systems a fractal Weyl law was found \cite{Sjo1990, Lin2002, LuSriZwo03,
SchTwo04, NonZwo05, KeaNovPraSie06, NonRub2007, WieMai2008, She2008,
RamPraBorFar2009, EbeMaiWun2010, ErmShe2010, PotEtAl2012, NonSjoZwo2013}.
Due to the opening of the system one classically obtains a fractal chaotic
saddle (sometimes also called repeller), which is the invariant set of points
in phase space that do not escape, neither in the future nor in the
past~\cite{LaiTel2011, AltPorTel2013}. 
Its fractal dimension $\delta$ plays an important role quantum mechanically:
The number $\mathcal{N}$ of long-lived resonance states is given by a fractal
Weyl law,
\begin{equation}\label{eq:FWL}
\mathcal{N} (h) \sim h^{-\delta/2} ,
\end{equation}
which here is stated for open chaotic maps, where the $k$ dependence is
replaced by the dependence on the effective size of Planck's cell $h$.

\begin{figure}[b]
\begin{center}
\includegraphics[scale=1.07]{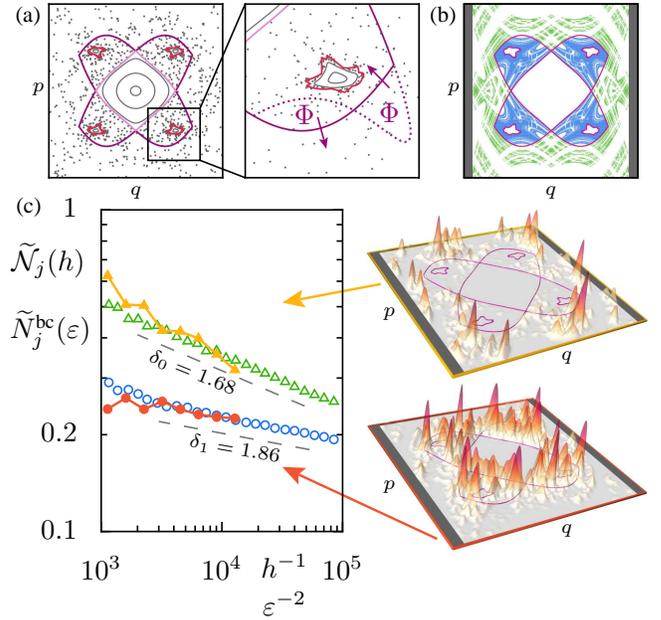}
\caption{
(Color online) (a) Phase space of the standard map at $\kappa =
2.9$ with regular (solid gray lines) and chaotic (gray points) orbits, three
partial barriers (solid colored lines) and the preimage of the outermost
partial barrier (dotted magenta line).
(b) Chaotic saddle of the opened map (gray-shaded absorbing stripes) colored
according to the regions ($A_0$: green, $A_1$: blue).
(c) Rescaled hierarchical fractal Weyl laws $\widetilde{\mathcal{N}}_j$ vs.\
$h^{-1}$ (filled symbols) counting hierarchical resonance states in the outer
($A_0$, triangles) and inner ($A_1$, circles) chaotic regions (with
corresponding typical Husimi representations for $h = 1/1000$). Their
power-law scaling is compared to the rescaled box-counting scaling
$\widetilde{N}^\text{bc}_j$ vs.\ $\varepsilon^{-2}$ (open symbols) with fractal
dimension $\delta_j$ in region $A_j$ of the chaotic saddle.
}
\label{fig:fig1}
\end{center}
\end{figure}

Generic Hamiltonian systems exhibit a mixed phase space where regular
and chaotic motion coexist~\cite{MarMey74}, see Fig.~\ref{fig:fig1}(a). Regular
resonance states of the open system obey a standard Weyl law, while for
chaotic resonance states one would naively expect that their number follows the
fractal Weyl law, Eq.~\eqref{eq:FWL}. This ignores, however, that the dynamics
in the chaotic region of generic two-dimensional maps is dominated by partial
transport barriers, see Fig.~\ref{fig:fig1}(a). 
A partial barrier is a curve which decomposes phase space into two
almost invariant regions. The small area enclosed by the partial barrier and
its preimage (dotted line in Fig.~\ref{fig:fig1}(a), magnification) consists of
two parts of size $\Phi$ on opposite sides of the partial barrier, which are
mapped to the other side in one iteration of the map. This flux $\Phi$ is the
characteristic property of a partial barrier.
There are infinitely many partial barriers which are hierarchically organized
with decreasing fluxes towards the regular regions~\cite{MacMeiPer84,
MacMeiPer84-2, HanCarMei85, MeiOtt85, Mei92}.
The partial barriers strongly impact the system's classical
\cite{MacMeiPer84, MacMeiPer84-2, HanCarMei85, MeiOtt85, Mei92, MotMouGreKan05,
CriKet08} and quantum mechanical~\cite{BroWya86, GeiRadRub1986, BohTomUll93,
Ket1996, SacEtAl1998, MaiHel00, KetHufSteWei00, BaeManHucKet02, MicEtAl2004,
GolGroTom09, MicBaeKetStoTom12} properties, and lead to, e.\,g., the
localization of eigenstates in phase space~\cite{BroWya86, GeiRadRub1986,
BohTomUll93, KetHufSteWei00, MicBaeKetStoTom12} and fractal conductance
fluctuations~\cite{Ket1996, SacEtAl1998, MicEtAl2004}.

Classically, the chaotic saddle, see Fig.~\ref{fig:fig1}(b), in generic
two-dimensional open maps gives rise to an individual fractal dimension for
each region of the hierarchical decomposition of phase
space~\cite{MotMouGreKan05}.
It is important to stress that these are \emph{effective} fractal dimensions,
which are constant over several orders, while in the limit of arbitrarily small
scales, they approach two~\cite{MotMouGreKan05, LauFinOtt91}.
Quantum mechanically, fractal Weyl laws for open systems with a mixed phase
space have been investigated in Refs.~\cite{KopSch10, SpiGarSar10,
IshAkaShuSch12, BirSch12}, but the influence of the hierarchical phase-space
structure remains open. In particular, the individual effective fractal
dimensions of the chaotic saddle have not been taken into account, so far.

In this Letter we propose a generalization of the Weyl law to open systems with
a mixed phase space. We obtain \emph{hierarchical fractal Weyl laws},
\begin{equation}\label{eq:indivFWL}
\mathcal{N}_j (h) \sim h^{-\delta_j/2} ,
\end{equation}
one for each phase-space region $A_j$ of the hierarchical decomposition of the
chaotic component in a generic two-dimensional phase space. Here, $\delta_j$
denotes the effective fractal dimension of the chaotic saddle in each region.
Quantum mechanically, this result is based on our observation of
\emph{hierarchical resonance states}, which predominantly localize on one of
the regions $A_j$. Their number $\mathcal{N}_j$ follows the hierarchical
fractal Weyl laws, Eq.~\eqref{eq:indivFWL}.
This holds over ranges of $h$ where on the corresponding classical scale the
effective fractal dimension $\delta_j$ is constant. In the semiclassical limit
we expect a scaling $h^{-1}$. Equation~\eqref{eq:indivFWL} is confirmed for the
generic standard map and a hierarchical model system.

\textit{Classical properties:}
We first review the classical properties of the chaotic saddle in a generic
mixed system and illustrate them for the prototypical example of the Chirikov
standard map~\cite{CasChiIzrFor79}. 
It is obtained from the kicked rotor Hamiltonian $H(q, p, t) = T(p) + V(q)
\sum_{n\in \mathbb{Z}} \delta(t - n)$ with kinetic energy $T(p) = p^2/2$ and
kick potential $V(q) = \frac{\kappa}{4\pi^2} \cos(2\pi q)$. At integer times
$t$ it leads to the symmetrized map $q_{t+1} = q_{t} + T'(p^*)$, $p_{t+1} = p^*
- \tfrac12 V'(q_{t+1})$ with $p^* = p_{t} - \tfrac12 V'(q_{t})$ on the torus
$[0, 1) \times \bigl[-\tfrac12, \tfrac12\bigr)$.
We open the system by defining absorbing stripes of width $0.05$ on the left
and right, see Fig.~\ref{fig:fig1}(b). 
This leads to a chaotic saddle $\Gamma$, for which a finite-time approximation
is shown in Fig.~\ref{fig:fig1}(b) for $\kappa =2.9$.
The chaotic saddle $\Gamma$ of the open system is strongly structured by 
the presence of partial barriers. They originate from Cantori or
stable/unstable manifolds of hyperbolic periodic orbits~\cite{Mei92}.
Partial barriers provide a hierarchical tree-like
decomposition~\cite{MeiOtt85}  of the chaotic component of phase space into
regions $A_j$:
A typical orbit explores a region $A_j$ before it enters a neighboring region
$A_k$. The transition rate is approximately given by the ratio $\Phi/A_j$ where
$\Phi$ is the flux across the partial barrier separating $A_j$ and $A_k$.
The route of escape from region $A_j$ to the opening is determined by the
tree-like decomposition of phase space. It traverses the sequence of
neighboring regions connecting $A_j$ with the opening in $A_0$.
The escape rate from a region $A_j$ is dominated by the first transition rate,
as subsequent transition rates are much larger.
Figure~\ref{fig:fig1}(a) shows the outermost partial barrier separating the
largest two regions $A_0$ and $A_1$ (which are quantum mechanically
accessible), as well as its preimage illustrating its flux $\Phi$. In addition
one can see the two partial barriers separating region $A_1$ from the chaotic
region near the central island and near the period-four regular island
chain. All three partial barriers are constructed from
stable/unstable manifolds of a period $4$ and a period $28$ orbit.

Using the box-counting method~\cite{Fal03} one can associate a fractal
dimension $\delta_j$ to the intersection $\Gamma \cap A_j$ of the chaotic
saddle $\Gamma$ with each of the regions $A_j$.
The number $N^\text{bc}_j(\varepsilon)$ of occupied boxes of side length
$\varepsilon$ scales like $N^\text{bc}_j(\varepsilon) \sim
\varepsilon^{-\delta_j}$, see Fig.~\ref{fig:fig1}(c), with $\delta_0 = 1.68$
and $\delta_1 = 1.86$.
To emphasize the difference between such dimensions close to two, the ordinate
is rescaled by $\varepsilon^{2}$, yielding the rescaled counting function
$\widetilde{N}^\text{bc}_j(\varepsilon) = N^\text{bc}_j(\varepsilon) \cdot
\varepsilon^2$.
The increase of $\delta_j$ towards two when going deeper into the hierarchy can
be qualitatively understood by adapting the Kantz--Grassberger
relation~\cite{KanGra85} from fully chaotic systems.

\begin{figure}
\begin{center}
\includegraphics[scale=1.062]{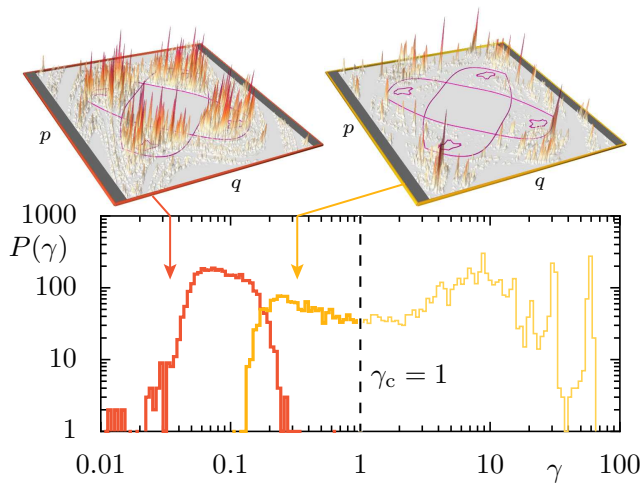}
\caption{
(Color online) Distributions $P(\gamma)$ of decay rates $\gamma$
for hierarchical resonance states of the standard map at $\kappa = 2.9$
located in regions $A_0$ (right, yellow) and $A_1$ (left, red) for $1/h =
12800$ and
corresponding Husimi representations of typical states. Short-lived
states ($\gamma > \gamma_\text{c}$) are not counted in the fractal Weyl law.
} \label{fig:fig2}
\end{center}
\end{figure}

\textit{Hierarchical resonance states:}
We now present the essential quantum effect that resonance states localize
predominantly on one of the regions $A_j$.
The closed quantum system is described by the time-evolution operator
$U=\exp\{-\frac{i}{2\hbar} V(q)\} \, \exp\{-\frac{i}{\hbar} T(p)\} \,
\exp\{-\frac{i}{2\hbar} V(q)\}$.
The corresponding open quantum system is given by $U_\text{open} = P U P$,
where $P$ is a projector on all positions not in the absorbing regions.
The resonance states $\psi$ are given by $U_\text{open} \psi = \exp[-i (\varphi
- i \gamma/2)] \psi$.
Regular resonance states are predominantly located in the regular region.
Chaotic resonance states are predominantly located in either of the
hierarchical regions $A_j$, see Fig.~\ref{fig:fig1}(c).
Hence, we will call them \textit{hierarchical resonance states} (of region
$A_j$).
Such a localization of chaotic eigenstates on different sides of a partial 
barrier is well known for closed quantum systems~\cite{MacMeiPer84,
BohTomUll93, MicBaeKetStoTom12}.
Chaotic eigenstates localized in the hierarchical region of a mixed phase space
were termed hierarchical states~\cite{KetHufSteWei00}.
They require that the classical flux $\Phi$ across a partial barrier is small
compared to the size $h$ of a Planck cell, i.\,e.\ $\Phi \ll h$. In the
opposite case, eigenstates would be equidistributed ignoring the partial
barrier~\cite{MacMeiPer84, BohTomUll93, MicBaeKetStoTom12}. 
Quite surprisingly, in open quantum systems we find that this condition
from closed systems is irrelevant for hierarchical resonance states.
In the standard map at $\kappa = 2.9$ we have $\Phi \approx 1/80$, and for $h
= 1/1000$, such that the condition $\Phi \ll h$ is violated, typical resonance
states still predominantly localize in one of the regions $A_j$, as shown in
Fig.~\ref{fig:fig1}(c). This is still the case for $h = 1/12800$, see
Fig.~\ref{fig:fig2}.
This crucial phenomenon for our study highlights the strong impact of the 
opening. 

One can qualitatively understand this localization of hierarchical resonance
states in the following way:
The localization on an almost invariant region $A_j$ seems plausible in view of
the semiclassical eigenfunction hypothesis for invariant
regions~\cite{Per1973, Ber1977, Vor1979}.
However, eigenstates localized on neighboring regions hybridize, if their
coupling due to the flux $\Phi$ is larger than their energy spacing. In closed
systems this happens for $\Phi > h$~\cite{MacMeiPer84, BohTomUll93,
MicBaeKetStoTom12}. In open systems, though, the distance of
resonance energies in the complex plane is larger due to their imaginary
part. In fact, it is much larger due to the different decay rates of
resonance states of neighboring regions $A_j$ corresponding to their different
classical escape rates. Therefore the localization of resonance states on
regions $A_j$ is possible in open systems, even if the criterion for the
closed system, $\Phi \ll h$, is not fulfilled. Such a line of reasoning is
reminiscent of the considerations on resonance trapping in fully chaotic
systems~\cite{PerGorRot1998, PerRot1999, PerRotStoBar2000}. This impact of the
opening will be studied quantitatively in the future.

For the present study it is sufficient to observe that the great majority of
chaotic resonance states is predominantly located in one of the regions
$A_j$, allowing their classification. Numerically, we use their relative local
Husimi weight in $A_j$ (in the case of $A_0$ excluding the area of the opening)
and discard states with more than 50\% Husimi weight in the regular region and
the deep hierarchical region ($A_j$, $j\geq 2$).
This classification is supported by the distribution of the decay rates
$\gamma$ of the corresponding resonance states, see Fig.~\ref{fig:fig2}.
States which are located deeper in the hierarchy have smaller decay rates.
In fact, the two distributions for regions $A_0$ and $A_1$ have a small
overlap, only.
Note that an alternative classification of resonance states purely based on
their decay rates $\gamma$, would fail deeper in the hierarchy, as the
tree-like structure allows for different regions $A_j$ having strongly
overlapping decay rate distributions.

\textit{Hierarchical fractal Weyl laws:}
For each region $A_j$ of the hierarchical phase space we now relate the number
$\mathcal{N}_j$ of hierarchical resonance states of that region to the fractal
dimension $\delta_j$ of the chaotic saddle in that region.
To this end we use the fractal Weyl law of fully chaotic systems
\cite{LuSriZwo03, SchTwo04}, Eq.~\eqref{eq:FWL}, individually for each region
$A_j$. This gives our main result that in open systems with a mixed phase space
one obtains a hierarchy of fractal Weyl laws, one for each phase-space region
$A_j$, Eq.~\eqref{eq:indivFWL}. We stress that this result is based on the
surprising existence of hierarchical resonance states.
Note that as a consequence of Eq.~\eqref{eq:indivFWL} the total number of
long-lived hierarchical resonance states is a superposition of power laws with
different exponents and not a single power law.

To give an intuitive understanding of the hierarchical fractal Weyl laws, let
us recall the interpretation of the fractal Weyl law~\cite{LuSriZwo03}, 
and apply it in the presence of a hierarchical phase space.
The number of quantum states localizing on a particular phase-space region
is given by the number of Planck cells necessary to cover the chaotic
saddle in that region.
Using the scaling, $N^\text{bc}_j(\varepsilon) \sim \varepsilon^{-\delta_j}$,
of the number of boxes $N^\text{bc}_j$ to cover the chaotic saddle in region
$A_j$ and the identification of the box area $\varepsilon^2$ with the Planck
cell area $h$ directly leads to Eq.~\eqref{eq:indivFWL}.
This holds for values of $h$ not too small, such that on the corresponding
classical scale the effective fractal dimension $\delta_j$ still remains
constant. Asymptotically ($\varepsilon \to 0$), all $\delta_j$ approach
two~\cite{MotMouGreKan05, LauFinOtt91}. 
Therefore, in the semiclassical limit ($h \to 0$), we expect an individual
resonance state to extend over all regions $A_j$ and that their number scales
as $h^{-1}$.

\textit{Standard map:}
The numerical investigation of the standard map supports the existence of
hierarchical fractal Weyl laws, as we now show. By the classification of
resonance states we are able to determine the number $\mathcal{N}_j(h)$ of
long-lived hierarchical resonance states associated with a particular
region $A_j$ depending on $h$. We restrict ourselves to the consideration of
small $h$ such that $\Phi /h \gtrsim 10$ where quantum mechanics can very well
mimic classical transport in phase space~\cite{MicBaeKetStoTom12}. 
Short-lived states are discarded by defining an arbitrary cut-off rate
$\gamma_\text{c} = 1$, as usual for the fractal Weyl law~\cite{SchTwo04}.
In globally chaotic systems the particular choice of $\gamma_\text{c}$
(within a reasonable range) does not influence the power-law exponent of the
fractal Weyl law but its prefactor only \cite{NonZwo05}. Here this
merely affects resonance states of the outermost region $A_0$. 
We obtain distinct behavior for each rescaled counting function
$\widetilde{\mathcal{N}}_j(h) = \mathcal{N}_j(h) \cdot h \cdot f_j$, see
Fig.~\ref{fig:fig1}(c), corresponding to the previous classical
rescaling. We fitted prefactors $f_j$ to the quantum results to better
demonstrate their scaling with power laws in agreement with the
classical counterparts (both prefactors $f_j$ are of order one: $f_0 = 2.6$,
$f_1 = 0.85$).
Apart from the smallest values of $1/h$, one observes the power-law scaling of
Eq.~\eqref{eq:indivFWL} and good agreement with the box-counting results for
the fractal dimensions $\delta_j$ of $\Gamma\cap A_j$. 
Note that the deviations between corresponding classical and quantum power-law
exponents are much smaller than the differences between the exponents
associated with different regions $A_j$ of the hierarchy.
Figure~\ref{fig:fig1}(c) confirms for two regions $A_j$ of the standard
map that they give rise to hierarchical fractal Weyl laws.
Note that the shape and position of the absorbing region modifies
the fractal dimension of the chaotic saddle and the power-law exponent of the
fractal Weyl law, but their relation remains valid (not shown).

\begin{figure}
\begin{center}
\includegraphics[scale=1.07]{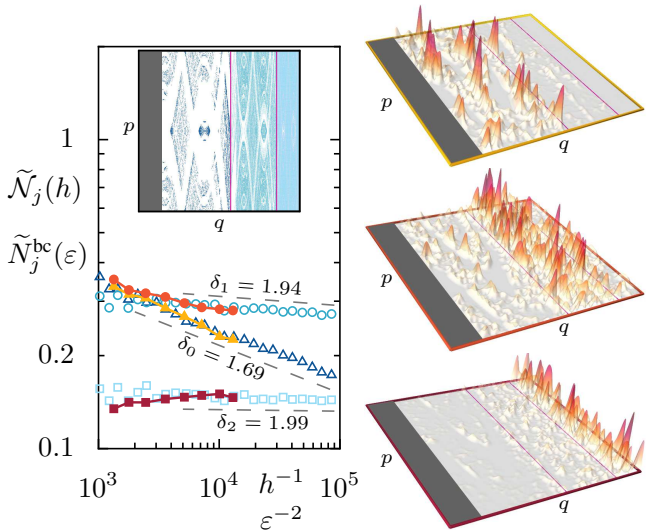}
\caption{
(Color online) Rescaled hierarchical fractal Weyl laws
$\widetilde{\mathcal{N}}_j$ vs.\ $h^{-1}$ (filled symbols) counting
hierarchical resonance states in the outer ($A_0$, triangles), central ($A_1$,
circles), and inner ($A_2$, squares) chaotic regions for the hierarchical
model system, and corresponding typical Husimi representations (right) for $h =
1/1115$. Comparison to rescaled box-counting scaling 
$\widetilde{N}^\text{bc}_j$ vs.\ $\varepsilon^{-2}$ (open symbols) of the
chaotic saddle (inset) with fractal dimension $\delta_j$ in region $A_j$.
} \label{fig:fig3}
\end{center}
\end{figure}

\textit{Hierarchical model system:}
To verify the hierarchical fractal Weyl laws for more than two regions, we
suggest the following system that models the hierarchical structure of partial
barriers in a generic mixed phase space, similar in spirit to a one-dimensional
model~\cite{MotMouGreKan05} and a Markov chain~\cite{HanCarMei85}.
The numerics for the corresponding quantum model allows for studying three
regions.

We first define a composed symplectic map $C \circ M$ on the phase space
$[0,1) \times [0, 1)$. It models $b$ partial barriers at the
positions $q_1 < \dots < q_b$ as straight lines in $p$-direction, giving a
decomposition into $b + 1$ regions $A_j = [ q_j, q_{j+1} ) \times [0,
1)$ with $q_0 = 0$ and $q_{b + 1} = 1$.
The map $M$ describes the uncoupled dynamics being sufficiently mixing in each
region $A_j$. We choose the standard map at kicking strength $\kappa = 10$
acting individually on each of the regions $A_j$ after appropriate rescaling.
The map $C$ couples these regions mimicking the turnstile mechanism of a
partial barrier with flux $\Phi_j$ by exchanging the areas $[q_j - \Phi_j, q_j)
\times [0, 1)$ with their neighboring areas $[q_j, q_j + \Phi_j) \times [0,
1)$. 
Finally, we open the system by defining the absorbing region $[0, \Phi_0)
\times [0, 1)$.
Here, we use $b = 2$, $q_1 = 4/7$, $q_2 = 6/7$, $\Phi_0 = 1/7$, $\Phi_1 =
1/28$, and $\Phi_2 = 1/112$.

Figure~\ref{fig:fig3} shows the results for the hierarchical model system: We
obtain the fractal dimensions $\delta_0 = 1.69$, $\delta_1 = 1.94$, and
$\delta_2 = 1.99$. Quantum mechanically, we again find hierarchical resonance
states predominantly localizing on one of the regions $A_j$, even though $h \ll
\Phi_1, \, \Phi_2$. Their number follows the proposed hierarchical fractal Weyl
laws according to Eq.~\eqref{eq:indivFWL}. For the rescaled numbers
$\widetilde{\mathcal{N}}_j$ in Fig.~\ref{fig:fig3} we use prefactors $f_0 =
1.75$, $f_1 = 1.55$, and $f_2 = 0.8$, which are of order one.

An experimental verification of the hierarchical fractal Weyl laws should be
feasible using microwave cavities as in a recent study on chaotic resonance
states~\cite{BarEtAl2013}.
A future challenge is the study of fractal Weyl laws in higher dimensional
systems with a generic phase space.

\begin{acknowledgments}
We are grateful to E.~G.~Altmann, H.~Kantz, H.~Schomerus, A.~Shudo, and    
T.~T\'el for helpful comments and stimulating discussions, and acknowledge
financial support through the DFG Forschergruppe 760 ``Scattering systems with
complex dynamics.''
\end{acknowledgments}

\end{document}